\documentclass[aps,prc,letterpaper,11pt,twoside,tightenlines,nofootinbib,showpacs,preprint]{revtex4}
\usepackage{graphicx}
\usepackage[sort&compress]{natbib}
\usepackage{subfigure}
\usepackage{amsmath}
\usepackage{amsfonts}
\usepackage{cancel}

\begin{document}

\arraycolsep1.5pt

\newcommand{\Ima}{\textrm{Im}}
\newcommand{\Rea}{\textrm{Re}}
\newcommand{\mev}{\textrm{ MeV}}
\newcommand{\be}{\begin{equation}}
\newcommand{\ee}{\end{equation}}
\newcommand{\ba}{\begin{eqnarray}}
\newcommand{\ea}{\end{eqnarray}}
\newcommand{\gev}{\textrm{ GeV}}
\newcommand{\nn}{{\nonumber}}
\newcommand{\dtres}{d^{\hspace{0.1mm} 3}\hspace{-0.5mm}}

\title{$NDK$, $\bar{K} DN$ and $ND\bar{D}$ molecules}

\author{C. W. Xiao $^1$, M. Bayar $^{1,\;2}$ and E. Oset $^1$}
\affiliation{$^1$ Departamento de F\'{\i}sica Te\'orica and IFIC, Centro Mixto Universidad de Valencia-CSIC,
Institutos de Investigaci\'on de Paterna, Aptdo. 22085, 46071 Valencia,
Spain\\
$^2$ Department of Physics, Kocaeli University, 41380 Izmit, Turkey}

\date{\today}

 \begin{abstract} 
 
We investigate theoretically baryon systems made of three hadrons which contain one nucleon and one D meson, and in addition another meson,  $\bar{D}, K$
or $\bar{K}$. The systems are studied using the Fixed Center Approximation to the Faddeev equations. The study is made assuming scattering of a $K$ or a $\bar{K}$ on a $DN$ cluster, which is known to generate the $\Lambda_c(2595)$, or the scattering of a nucleon on the $D\bar{D}$ cluster, which has been shown to generate a hidden charm resonance named X(3700). We also investigate the configuration of scattering of $N$ on the $KD$ cluster, which is known to generate the $D_{s0}^*(2317)$. In all cases we find bound states, with the $NDK$ system, of exotic nature, more bound than the $\bar{K} DN$.
\end{abstract}


\maketitle

\section{Introduction}
\label{Intro}
    While the three baryon system has been a subject of intense theoretical study \cite{Sauer:2010zz,Epelbaum:2008zz,Hiyama:2006zz}, it has only been recently that attention was brought to systems with two mesons and one baryon, with unexpected results. Indeed, states with two pseudoscalar mesons and one baryon were studied in \cite{alberone}, combining Faddeev equations and chiral dynamics, by means of which the low lying excited $J^P=1/2^+$ $\Lambda$ and $\Sigma$ states of the Particle Data Book (PDG) \cite{pdg} were described. The same happened with the low lying  $J^P=1/2^+$ $N^*$ states \cite{alberdos}. Independently, and using variational techniques, a $N^*$ state around 1920 MeV 
was predicted in \cite{jidoenyo} as a molecule of $NK \bar{K}$, corroborated in coupled channels Faddeev equations in \cite{nstarpheno} and  \cite{MartinezTorres:2010zv}. A different calculation also predicts a quasibound $\pi \bar{K} N$ system, with the difference that the $N \pi$ interaction is in p-wave \cite{garcigal}.  Systems of bound or quasibound three mesons did not wait long and in \cite{albermeson} the X(2175) (now the $\phi(2170)$) was explained as a resonant $\phi K \bar{K}$ system, also described as such in a  phenomenological way in \cite{AlvarezRuso:2009xn}. In a similar way, the  $K(1460)$ is explained as a $K K \bar{K}$ state in \cite{alberjido} and more recently the $\pi (1300)$ is described as a $\pi K \bar{K}$ molecule in \cite{albernew}.

    The charm sector has not yet been explored for such three body systems and this is the first incursion in that field. For this purpose we have selected systems that have a nucleon and a $D$ meson. The $DN$ system, in collaboration with coupled channels, leads to the formation of a dynamically generated state, the 
$\Lambda_c(2595)$ \cite{lutzcharm, mizuangels,conlaura,conjuan}. On top of it we add a $K,\bar{K}$ or $\bar{D}$ meson and we study the stability of the system. The case of scattering of $N$ on the $DK$ cluster, which is known to generate the $D_{s0}^*(2317)$ \cite{Hofmann:2003je,Guo:2006fu,danielfirst} is also considered. 
On the other hand, the $D \bar{D}$ system leads to a bound state in isospin I=0 \cite{danielfirst}, which might have already been observed \cite{Gamermann:2007mu}, in analogy to the $f_0(980)$ which is largely made of $K \bar{K}$ \cite{Isgur,npa,norbert,prl,ramonet,hanhart1,hanhart2}. We add a nucleon to it and study the interaction of the three body system. The system obtained would be the analogous to the one found in \cite{jidoenyo,nstarpheno,MartinezTorres:2010zv} as a $N K \bar{K}$. In all cases we find bound or quasibound states with masses around 3100 MeV in the first cases and 4400 MeV in the case of $N D \bar{D}$.

  The calculations are made using the Fixed Center Approximation (FCA) to the Faddeev equations. The method has proved to be rather reliable for cases like $K$-deuteron scattering very close to threshold (see \cite{Gal:2006cw} and the earlier work \cite{Toker:1981zh}). In a closer problem to the present one, diverting a bit from threshold in the bound region, the FCA has been applied to the study of the $N K \bar{K}$ system in \cite{Xie2010} and the results compare favorably with those of the Faddeev approach in \cite{nstarpheno} and those of the variational approach in \cite{jidoenyo}. Similarly the FCA has been applied to an analogous problem, the one of the $\bar{K} NN$ system \cite{Bayar:2011qj} and a detailed exposition has been made in the Introduction of the approximations involved in ordinary Faddeev calculations, which can induce uncertainties as large as those of the FCA for different reasons. Yet, it has been reassuming to see in that paper that the results of the FCA are qualitatively in agreement with those of other approaches \cite{Ikeda:2007nz,Shevchenko:2006xy,Shevchenko:2007zz,Dote:2008in,Dote:2008hw,Ikeda:2008ub}, and remarkably similar to those obtained in the variational approach of \cite{Dote:2008in,Dote:2008hw} which use the same input as in \cite{Bayar:2011qj}. The differences between \cite{Dote:2008in,Dote:2008hw} and \cite{Bayar:2011qj} are at the level of a few MeV in the binding, an accuracy which is far more that sufficient in the exploratory work that we carry on the systems under consideration, which are studied here for the first time.

\section{Multi-body interaction formalism}
\label{formalism}
   The Faddeev equations under the FCA are an effective tool to deal with multi-hadron interaction \cite{Xie2010,Bayar:2011qj,Roca2010,Junko2010,Xie2011}. They are particularly suited to study system in which a pair of particles cluster together and the cluster is not much modified by the third particle. Even if there is a sizeable modification of the cluster, the method is useful in combination with information from other sources on how the cluster can be modified by the presence of the third particle \cite{Bayar:2011qj}. The assumption of a small modification of the cluster wave function seems to imply that the mass of the third particle should be smaller than that of the cluster components, but this can also happen if the cluster is strongly bound, independent of the masses. In any case, in an exploratory study of the systems under consideration, where uncertainties of even $50\mev$ are readily acceptable, the FCA proves to be a sufficiently accurate tool, and relatively simple to use, once comparison with more accurate tools has shown that the same results are obtained within a few MeV of difference.
  
\begin{figure}
\centering
\includegraphics[scale=0.4]{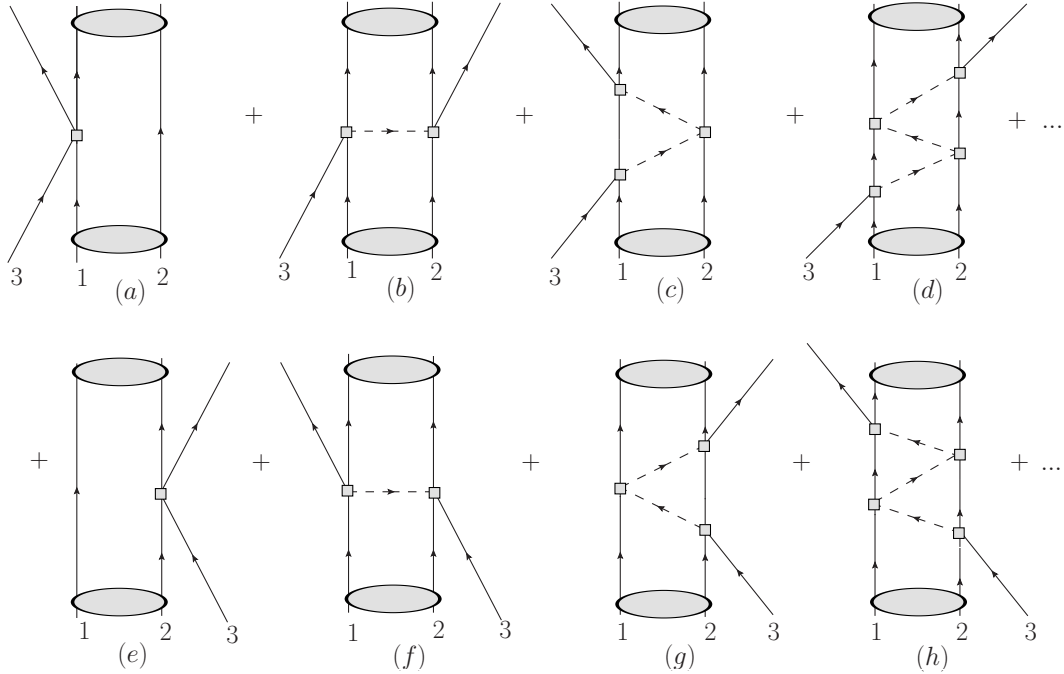}
\caption{Diagrammatic representation of the FCA to Faddeev equations.}\label{FCAfig}
\end{figure}

   Following \cite{Xie2010,Bayar:2011qj,Roca2010,Junko2010,Xie2011}, we will apply the FCA to study the charm sector. The FCA approximation to Faddeev equations assumes a pair of particles (1 and 2) forming a cluster. Then particle 3 interacts with the components of the cluster, undergoing all possible multiple scattering with those components. This is depicted in Fig. \ref{FCAfig}.
With this meaning of the FCA, it is easy to write the equations. For this one defines two partition functions $T_1$, $T_2$ which sum all diagrams of the series of Fig. \ref{FCAfig} which begin with the interaction of particle 3 with particle 1 of the cluster ($T_1$), or with the particle 2 ($T_2$). The equations then read 
\begin{align}
T_1&=t_1+t_1G_0T_2,\label{threet1}\\
T_2&=t_2+t_2G_0T_1,\label{threet2}\\
T&=T_1+T_2,\label{threet}
\end{align}
where $T$ is the total three-body scattering amplitude that we are looking for. The amplitudes $t_1$ and $t_2$ represent the unitary scattering amplitudes with coupled channels for the interactions of particle 3 with particle 1 and 2, respectively. Besides, $G_0$ is the propagator of particle 3 between the components of the two-body system, which will be discussed below.
  
   Now we turn to the amplitudes corresponding to single-scattering contribution. One must take into account the isospin structure of the cluster and write the $t_1$ and $t_2$ amplitudes in terms of the isospin amplitudes of the (3,1) and (3,2) systems. Details can be seen in \cite{Junko2010,Xie2010}. In the present case this is particularly easy. Whether we have the $\bar{K}DN,~KDN,~NDK$ or $ND\bar{D}$ system, where the first particle is labelled 3 and the last two particles are making the cluster (particles 1, 2), the bound state of 1, 2 is in $I=0$ from former studies and the total spin is then $I=\frac{1}{2}$. Then, for all four cases we find
\begin{align}
t_1&=\frac{3}{4}~ t_{31}^{I=1} +\frac{1}{4}~ t_{31}^{I=0},\label{t1}\\
t_2&=\frac{3}{4}~ t_{32}^{I=1} +\frac{1}{4}~ t_{32}^{I=0},\label{t2}
\end{align}
in which $t_{31}^{I=1}$ is the two-body unitary scattering amplitude of isospin $I=1$ between particle 3 and 1 evaluated with its coupled channels, and similarly for the other cases. We show below the explicit evaluation for the $ND \bar{D}$ case. The derivation for the other systems is identical. Here we take the case of $I_{D \bar{D}}=0$ and $I_{total}\equiv I_{ND \bar{D}}=1/2$. We have
\be 
|D \bar{D}>^{(0,0)}=\sqrt{\frac{1}{2}}|(\frac{1}{2},-\frac{1}{2})>-\sqrt{\frac{1}{2}}|(-\frac{1}{2},\frac{1}{2})>,
\ee
where $|(\frac{1}{2},-\frac{1}{2})>$ denote $|(I_z^1,I_z^2)>$ which shows the $I_z$ components of particles 1 and 2. Then we obtain
\be 
\begin{split}
t=&<ND \bar{D}|\,\hat{t}\,|ND \bar{D}>\\
=&(<D \bar{D}|^{(0,0)} \otimes <N|^{(\frac{1}{2},\frac{1}{2})})\,(\hat{t}_{31}+\hat{t}_{32})\,(|D \bar{D}>^{(0,0)} \otimes |N>^{(\frac{1}{2},\frac{1}{2})})\\
=&\bigg[\sqrt{\frac{1}{2}}\bigg(<(\frac{1}{2},-\frac{1}{2})|-<(-\frac{1}{2},\frac{1}{2})|\bigg) \otimes <N|^{(\frac{1}{2},\frac{1}{2})}\bigg]\,(\hat{t}_{31}+\hat{t}_{32})\,\bigg[\sqrt{\frac{1}{2}}\bigg(|(\frac{1}{2},-\frac{1}{2})>\\
&-|(-\frac{1}{2},\frac{1}{2})>\bigg) \otimes |N>^{(\frac{1}{2},\frac{1}{2})}\bigg]\\
=&\bigg[\sqrt{\frac{1}{2}}<(1,1),-\frac{1}{2}|-\sqrt{\frac{1}{2}}\bigg(\sqrt{\frac{1}{2}}<(1,0),\frac{1}{2}|-\sqrt{\frac{1}{2}}<(0,0),\frac{1}{2}|\bigg)\bigg]\,\hat{t}_{31}\,\bigg[\sqrt{\frac{1}{2}}|(1,1),-\frac{1}{2}>\\
&-\sqrt{\frac{1}{2}}\bigg(\sqrt{\frac{1}{2}}|(1,0),\frac{1}{2}>-\sqrt{\frac{1}{2}}|(0,0),\frac{1}{2}>\bigg)\bigg]
+\bigg[\sqrt{\frac{1}{2}}\bigg(\sqrt{\frac{1}{2}}<(1,0),\frac{1}{2}|-\sqrt{\frac{1}{2}}<(0,0),\frac{1}{2}|\bigg)\\
&-\sqrt{\frac{1}{2}}<(1,1),-\frac{1}{2}|\bigg]\,\hat{t}_{32}\,\bigg[\sqrt{\frac{1}{2}}\bigg(\sqrt{\frac{1}{2}}|(1,0),\frac{1}{2}>-\sqrt{\frac{1}{2}}|(0,0),\frac{1}{2}>\bigg)-\sqrt{\frac{1}{2}}|(1,1),-\frac{1}{2}>\bigg]\\
=&\Big(\frac{3}{4}t_{ND}^{I=1}+\frac{1}{4}t_{ND}^{I=0}\Big)+\Big(\frac{3}{4}t_{N \bar{D}}^{I=1}+\frac{1}{4}t_{N \bar{D}}^{I=0}\Big),\\
\end{split}
\ee
where the notation of the states followed in the terms is $|(1,1),-\frac{1}{2}> \equiv |(I^{31},I_z^{31}),I_z^2>$ for $t_{31}$, and $|(I^{32},I_z^{32}),I_z^1>$ for $t_{32}$.

  Because we follow the normalization of Mandl and Shaw \cite{mandl} which has different weight factors for the meson and baryon fields, we must take into account  how these factors appear in the single scattering and double scattering and in the total amplitude. This is easy and is done in detail in \cite{Junko2010,Xie2010} for the two different cases, when the cluster is a baryon ($KDN,~\bar{K}DN$) or a meson ($ND\bar{D},~NDK$). We show below the details for the case of a baryonic cluster and a meson as scattering particle. Let us assume that particle 1 of the cluster is a meson and particle 2 is a baryon. Particle 3 is also a meson. In this case, following the field normalization of \cite{mandl} we find for the S matrix of single scattering,
\begin{align}
\begin{split}
S^{(1)}_1=&-it_1 \frac{1}{{\cal V}^2} \frac{1}{\sqrt{2\omega_3}} \frac{1}{\sqrt{2\omega'_3}}
 \frac{1}{\sqrt{2\omega_1}} \frac{1}{\sqrt{2\omega'_1}}\\
 &\times(2\pi)^4\,\delta(k+k_R-k'-k'_R),\\
\end{split}\\
\begin{split}
S^{(1)}_2=&-it_2 \frac{1}{{\cal V}^2} \frac{1}{\sqrt{2\omega_3}} \frac{1}{\sqrt{2\omega'_3}}
 \sqrt{\frac{2M_2}{2E_2}} \sqrt{\frac{2M_2}{2E'_2}}\\
 &\times(2\pi)^4\,\delta(k+k_R-k'-k'_R),\\
\end{split}
\end{align}
where, $k,\,k'$ ($k_R,\,k'_R$) refer to the momentum of initial, final scattering particle (R for the cluster), $\cal V$ is the volume of the box where the states are normalized to unity and the subscripts 1, 2 refer to scattering with particle 1 or 2 of the cluster.

  The double scattering diagram, Fig. \ref{FCAfig} (b), is given by
\be 
\begin{split}
S^{(2)}=&-i(2\pi)^4 \delta(k+k_R-k'-k'_R) \frac{1}{{\cal V}^2}
\frac{1}{\sqrt{2\omega_3}} \frac{1}{\sqrt{2\omega'_3}}
 \frac{1}{\sqrt{2\omega_1}} \frac{1}{\sqrt{2\omega'_1}}
 \sqrt{\frac{2M_2}{2E_2}} \sqrt{\frac{2M_2}{2E'_2}}\\
&\times\int \frac{d^3q}{(2\pi)^3} F_R(q) \frac{1}{{q^0}^2-\vec{q}\,^2-m_3^2+i\,\epsilon} t_{1} t_{2},
\end{split} 
\ee 
where $F_R(q)$ is the cluster form factor that we shall discuss below.

  Similarly the full S matrix for scattering of particle 3 with the cluster will be given by
\be 
\begin{split}
S=&-i\, T \, (2\pi)^4 \delta(k+k_R-k'-k'_R)\frac{1}{{\cal V}^2}\\
&\times\frac{1}{\sqrt{2 \omega_3}} \frac{1}{\sqrt{2 \omega'_3}}
\sqrt{\frac{2M_R}{2E_R}} \sqrt{\frac{2M_R}{2E'_R}}.
\end{split} 
\ee
In view of the different normalization of these terms, we can introduce suitable factors in the elementary amplitudes,
\be
\tilde{t_1}=\frac{1}{2m_1}~ t_1,~~~~\tilde{t_2}=t_2,
\ee
where we have taken the approximations, suitable for bound states, $\frac{1}{\sqrt{2 \omega_i}}=\frac{1}{\sqrt{2m_i}},\, \sqrt{\frac{2M_R}{2E_R}}=1$, and sum all the diagrams by means of
\be 
T=\frac{\tilde{t_1}+\tilde{t_2}+2~\tilde{t_1}~\tilde{t_2}~G_0}{1-\tilde{t_1}~\tilde{t_2}~G_0^2}, \label{new}
\ee
The function $G_0$ in Eq. \eqref{new} is given by 
\be 
G_0(s)=\int \frac{d^3\vec{q}}{(2\pi)^3} F_R(q) \frac{1}{q^{02}-\vec{q}^{~2}-m_3^2 +i\,\epsilon }.\label{g0baryon}
\ee
where $F_R(q)$ is the form factor of the cluster of particles 1 and 2. We must use the form factor of the cluster consistently with the theory used to generate the cluster as a dynamically generated  resonance. This requires to extend into wave functions the formalism of the chiral unitary approach developed for scattering amplitudes \cite{Kaiser:1995eg,angels,ollerulf,Lutz:2001yb,osetplb527,Hyodo:2002pk,cola, Borasoy:2005ie,Oller:2006jw,Borasoy:2006sr}. This work has been done recently in \cite{gamerjuan,yamajuan} and the expression for the form factors are given in section 4 of \cite{yamajuan}, which we use in the present work and reproduce below 
\begin{align}
\begin{split}
F_R(q)&=\frac{1}{\mathcal{N}} \int_{|\vec{p}|<\Lambda, |\vec{p}-\vec{q}|<\Lambda} d^3 \vec{p} \frac{1}{2 E_1(\vec{p})} \frac{1}{2 E_2(\vec{p})} \frac{1}{M_R-E_1(\vec{p})-E_2(\vec{p})} \\
&\quad\frac{1}{2 E_1(\vec{p}-\vec{q})} \frac{1}{2 E_2(\vec{p}-\vec{q})} \frac{1}{M_R-E_1(\vec{p}-\vec{q})-E_2(\vec{p}-\vec{q})}, \label{formfactor}
\end{split}\\
\mathcal{N}&=\int_{|\vec{p}|<\Lambda} d^3 \vec{p} (\frac{1}{2 E_1(\vec{p})} \frac{1}{2 E_2(\vec{p})} \frac{1}{M_R-E_1(\vec{p})-E_2(\vec{p})})^2, \label{formfactorN}
\end{align}
where $E_1$ and $E_2$ are the energies of the particles 1, 2 and $M_R$ the mass of the cluster. The parameter $\Lambda$ is a cut off that regularizes the integral of Eqs. \eqref{formfactor} and \eqref{formfactorN}. This cut off is the same one needed in the regularization of the loop function of the two particle propagators in the study of the interaction of the two particles of the cluster \cite{yamajuan}. We take these values of $\Lambda$ such as to get the bound states of $\Lambda _c(2595)$ from \cite{mizuangels}, $D_{s0}^*(2317)$ from \cite{danielfirst} and $X(3700)$ from \cite{danielfirst}.

  When the cluster is a meson and the scattering particle a baryon, the solution is given by the same Eq. \eqref{new}, but now
\begin{align}
\tilde{t_1}&=\frac{2m_R}{2m_1}~ t_1,~~~~\tilde{t_2}=\frac{2m_R}{2m_2}~ t_2,\\
G_0(s)&=\frac{1}{2m_R}\int \frac{d^3\vec{q}}{(2\pi)^3} F_R(q) \frac{m_3}{E_3(\vec{q})}\frac{1}{q^{0} -E_3(\vec{q}) + i\,\epsilon } . \label{g0meson}
\end{align}  
In addition, $q^0$, the energy carried by particle 3 in the rest frame of the three particle system, is given by 
\be 
q^0(s)=\frac{s+m_3^2-M_R^2}{2\sqrt{s}}.
\ee  

   Note also that the arguments of the amplitudes $T_i(s)$ and $t_i(s_i)$ are different, where $s$ is the total invariant mass of the three-body system, and $s_i$ are the invariant masses in the two-body systems. The value of $s_i$ is given by \cite{Junko2010}
\be 
s_i=m_3^2+m_i^2+\frac{(M_R^2+m_i^2-m_j^2)(s-m_3^2-M_R^2)}{2M_R^2}, (i,j=1,2,\;i\neq j)
\ee
where $m_l, (l=1,2,3)$ are the masses of the corresponding particles in the three-body system and $M_R$ the mass of two body resonance or bound state (cluster).

\section{The case of $\bar{K}DN$ interaction}
\label{kbdn}
\begin{figure}
\centering
\includegraphics[scale=1]{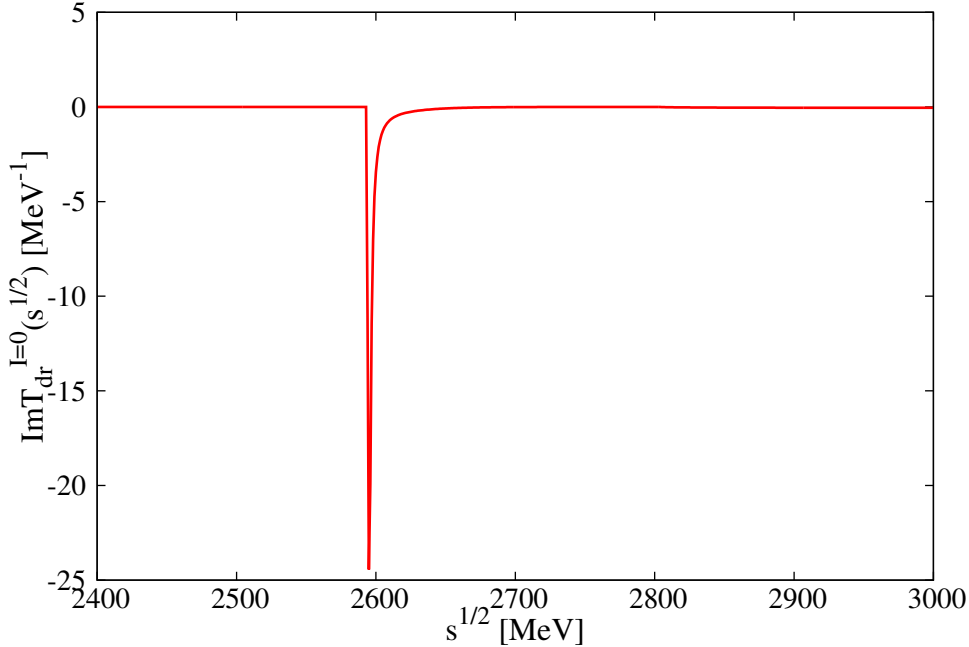}
\caption{Imaginary part of the $DN$ amplitude for isospin $I=0$.}\label{fig2}
\end{figure}

\begin{figure}
\centering
\includegraphics[scale=1]{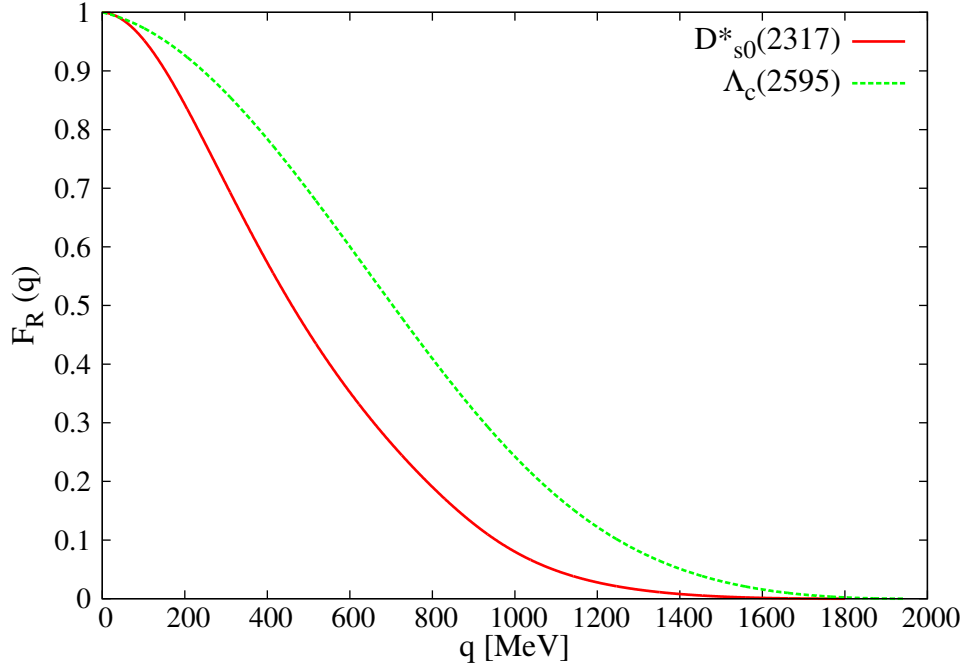}
\caption{Form factors of $\Lambda _c(2595)$ and $D_{s0}^*(2317)$.}\label{fig1}
\end{figure}

\begin{figure}
\centering
\includegraphics[scale=1]{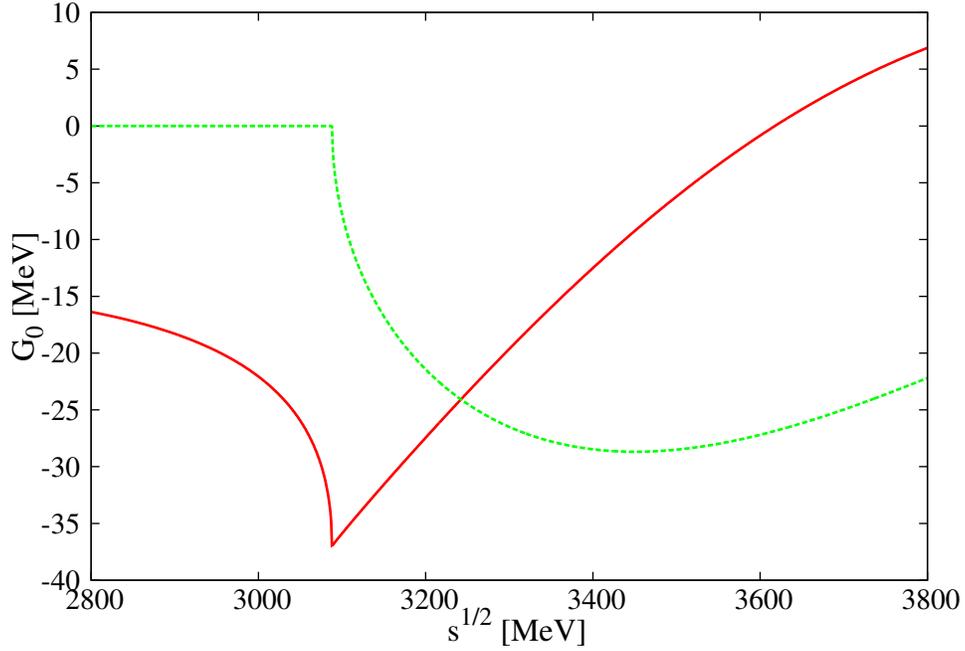}
\caption{Real(solid line) and imaginary(dashed line) parts of $G_0$ in $\bar{K}DN$.}\label{fig3}
\end{figure}

   Our strategy proceeds as follows: first we generate the resonance or bound state in the compound system and determine the value of the parameter $\Lambda$, then calculate the form factor and $G_0$ propagator and take the $t_1$ and $t_2$ amplitudes from the unitary coupled channel approach, finally the total scattering amplitude $T$ is evaluated. For the $\bar{K}DN$ scattering, the first thing we do is to reproduce the work of \cite{mizuangels,Tolos:2009qz} in coupled channels for the $DN$ system. The coupled channels used are $\pi \Sigma_c,\, DN,\, \eta \Lambda_c,\, K\Xi_c,\, K\Xi'_c,\, D_s \Lambda,\, \eta'\Lambda_c$ and the dynamics is obtained from the exchange of vector mesons between the pseudoscalar meson and the baryon. This procedure, based on the local hidden gauge approach \cite{hidden1,hidden2,hidden3}, leads to the chiral Lagrangians in the SU(3) sector.
    One gets as dynamically generated resonance the $\Lambda _c(2595)$, which couples most strongly to the $DN$ channel and is interpreted as a $DN$ bound state. As shown in Fig. \ref{fig2}, the method generates a pole in the $DN$ scattering amplitude in $I=0$ in the first Riemann Sheet at $(2595+i0)\mev$. Since the works of \cite{mizuangels,Tolos:2009qz} use dimensional regularization for the loops, and we need a cut off to obtain the wave function and form factor, the equivalent cut off must be obtained. There are three methods to do this. One of them is to compare the value of the $G$ propagator (loop function of two particles propagator which appears in the Bethe-Salpter equation $T=V+VGT$)
     at threshold using the dimensional regularization formula \cite{osetplb527} with the one of the cut off which can be taken from \cite{angels} (\cite{npa} for meson-meson interaction) or the analytic expression in \cite{Guo2006}. Another method is to compare the pole position using different regularization schemes. The third one is to use the relation between the subtraction constant $a(\mu)$ and the cut off $\Lambda$ of Eq. (52) in \cite{GarciaRecio:2010ki}. The best fitting results by these methods give us a value of $\Lambda=973\mev$. The $\Lambda_c(2595)$ form factor using this cut off is shown in Fig. \ref{fig1}. In the next step we evaluate $G_0$ by means of Eq. \eqref{g0baryon} and we show its real and imaginary parts in Fig. \ref{fig3}.  

    In the final calculation we also need to know the two-body unitary scattering amplitudes in different isospin states. For the $\bar{K}DN$ interaction, the amplitudes of $\bar{K}D$ and $\bar{K}N$ are needed. For the $\bar{K}D$ interaction we have taken the results of \cite{danielfirst}. On the other hand, the $\bar{K}N$ scattering has been evaluated using the chiral unitary approach of \cite{angels} with the dimensional regularization scheme of \cite{osetplb527}, using $\mu=630\mev, a_i(\mu)=-1.84$ for all channels. This scheme leads to the generation of the two $\Lambda(1405)$ states reported in \cite{cola}.
    
\begin{figure}
\centering
\includegraphics[scale=1]{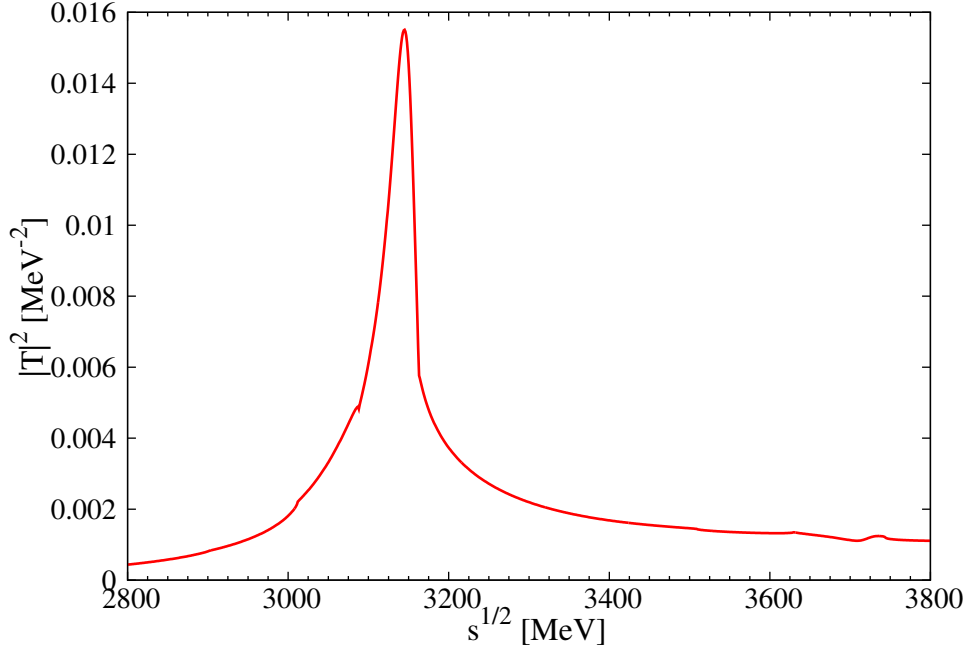}
\caption{Modulus squared of the $\bar{K}\Lambda_c(2595)$ scattering amplitude.}\label{fig4}
\end{figure}

   In Fig. \ref{fig4} we show the results of $|T|^2$ for the $\bar{K}\Lambda_c(2595)$ scattering. We find a peak around $3150\mev$, slightly above the threshold of the $\Lambda_c(2595)+\bar{K}$ mass ($3088\mev$) and below the threshold of the $\bar{K}DN$ system ($3298\mev$). The width of the peak is about $50\mev$, which indicates the width of the state that we obtain. In our study of the system $\bar{K}DN$, where we have chosen the $DN$ system in $I=0$ since this is the channel with strong attraction leading to the $\Lambda_c(2595)$ resonance, the quantum numbers of the $\bar{K}DN$ state are $C=+1,S=-1$ and $J^P=\frac{1}{2}^+$ since we only consider the interaction among the components in $L=0$. The mass of this state is very close to that of the $\Xi_c(3123)$, of unknown spin and parity, but its decay width is larger than that of the $\Xi_c(3123)$, $4\pm 4\mev$ \cite{pdg}. The larger width, tied to the $\pi \Sigma$ decay of the $\bar{K} N$ system, seems unavoidable, and this indicates that the resonance that we find could most probably correspond to a resonance not yet found.

\section{Investigating the $NDK$ interaction}
\label{InveNDK}
   The two body $KD$ and $DN$ interactions were studied by the coupled channel Bethe-Salpeter equations  in \cite{danielfirst} and \cite{mizuangels,Tolos:2009qz}, respectively.
 It was found that the resonance $D_{s0}^*(2317)$ is dynamically generated in $I=0$ from $KD$ 
scattering and the $\Lambda _c(2595)$ is produced in $I=0$ from the $DN$ interaction, as we mentioned above. Hence there are two possible cases of 
three body scattering in the $NDK$ system. One is the $N-(DK)_{D_{s0}^*(2317)}$ and the other one is the $K-(DN)_{\Lambda_c(2595)}$.

\begin{figure}
\centering
\includegraphics[scale=1]{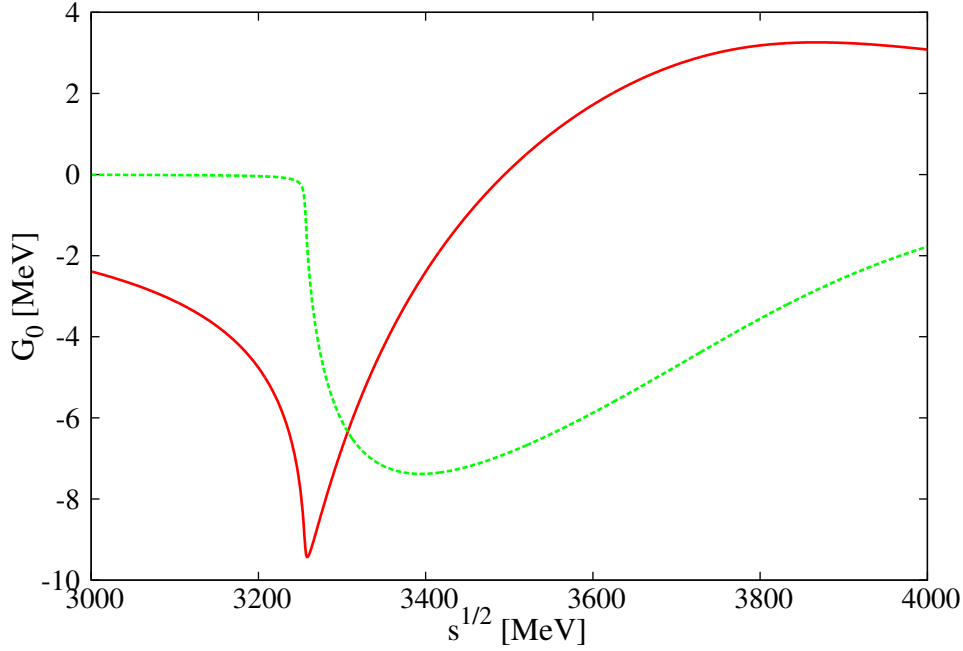}
\caption{Real(solid line) and imaginary(dashed line) parts of $G_0$ in $N-DK$.}\label{gndk}
\end{figure}

First, we are going to investigate the three body scattering for the $N-(DK)_{D_{s0}^*(2317)}$. In order to calculate this, in a first step
we obtain the $DN$ amplitude, $t_{1}$ from \cite{mizuangels, Tolos:2009qz} and the $KN$  amplitude, $t_{2}$ from the chiral unitary approach of
 \cite{angels}. For the $DN$ matrix element  the
 result of \cite{mizuangels} is reproduced and the imaginary part of the $DN$ matrix element that we obtain is shown in Fig. \ref{fig2}.
 In the case of the  $KN$ system, the interaction is repulsive in $I=1$ and has vanishing interaction in $I=0$. 
We take the parameters for the loop function from \cite{osetplb527}.
For the next step, in order to get the total scattering amplitude $T$, we need to know the form factor for $D_{s0}^*(2317)$ and $G_0(s)$. Here the cut off 
is determined by comparing the value of the G function that one obtains from the dimensional regularization \cite{osetplb527} and
the cut off scheme \cite{angels} at threshold. In this way $\Lambda=900\mev$ is obtained for the cut off of the $D_{s0}^*(2317)$ form factor. 
The form factor of the $D_{s0}^*(2317)$ is plotted in Fig. \ref{fig1} and the $G_0(s)$ function for this case is represented in Fig. \ref{gndk}.

\begin{figure}
\centering
\includegraphics[scale=1]{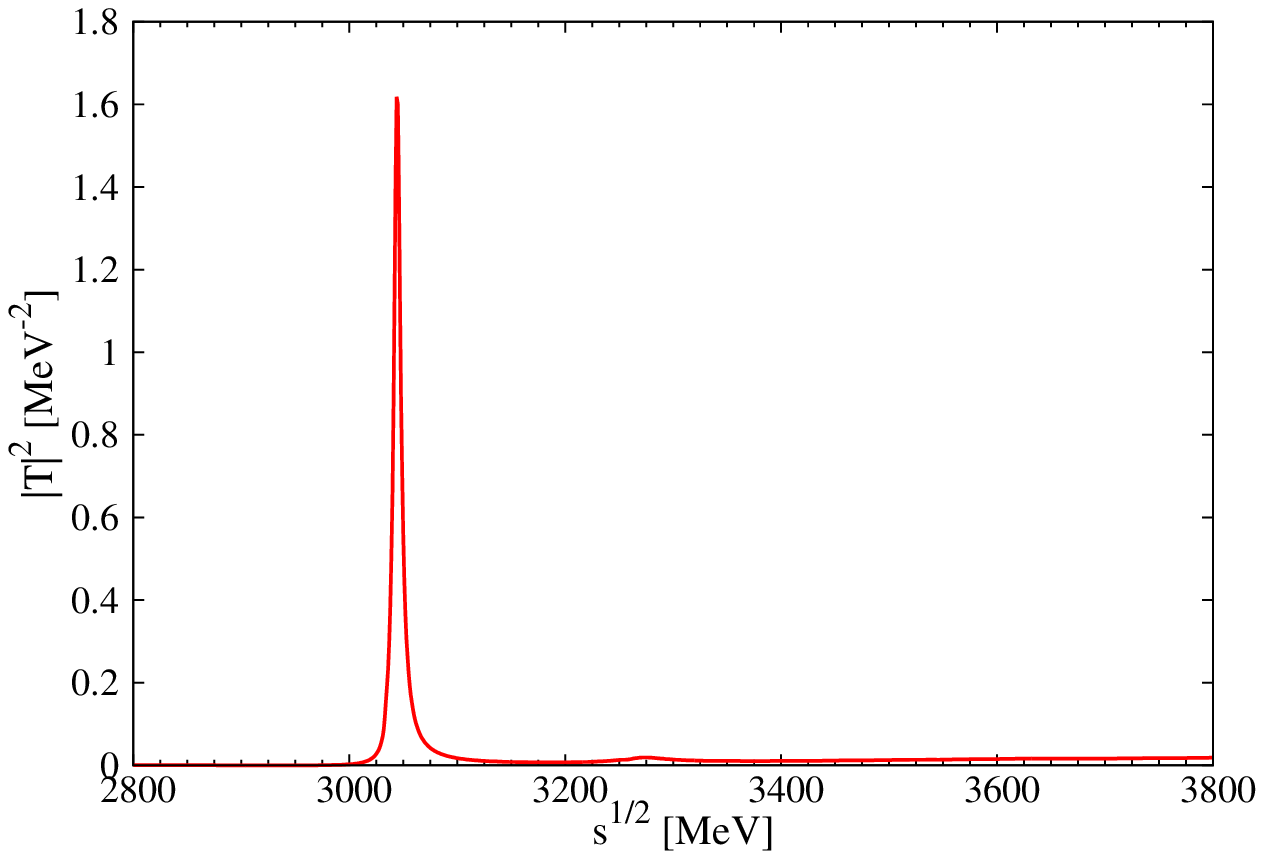}
\caption{Modulus squared of the $ND_{s0}^*(2317)$ scattering amplitude.}\label{tndk}
\end{figure}

Using the aforementioned total three-body scattering amplitude $T$, we obtain $|T|^2$ for the  $ND_{s0}^*(2317)$ scattering shown 
in Fig. \ref{tndk}. We found a peak around $3050\mev$ which is about $200\mev$ below the $D_{s0}^*(2317)$ and $N$ threshold. This reflects
the strong attraction in the $DN$ system that leads to  the $\Lambda _c(2595)$. The width of the state is smaller than $10\mev$. We do not 
find a counterpart in the PDG and the quantum numbers, with positive strangeness, correspond to an exotic state.

\begin{figure}
\centering
\includegraphics[scale=1]{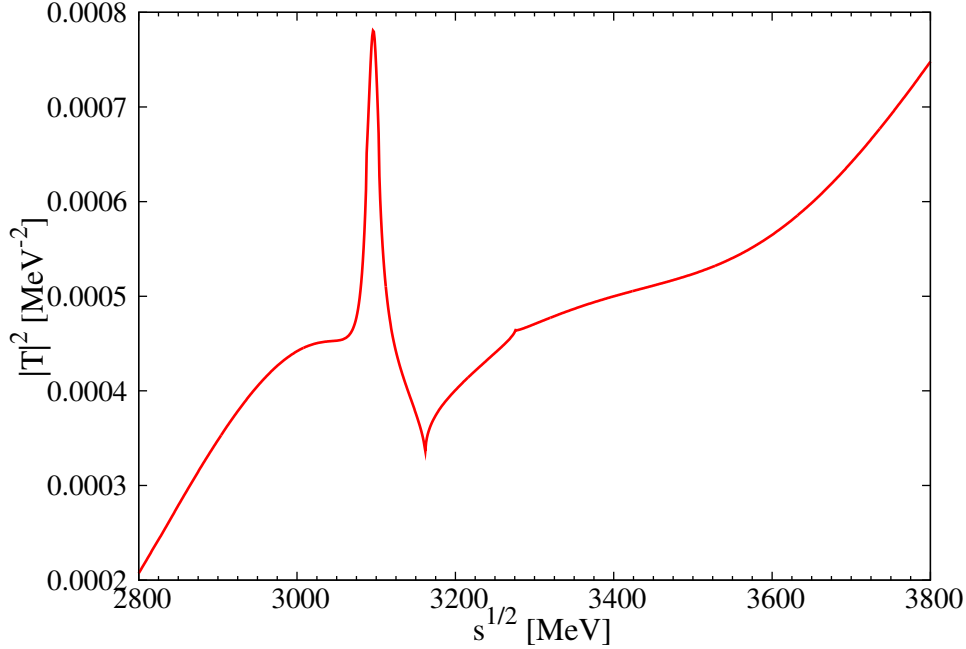}
\caption{Modulus squared of the $K\Lambda_c(2595)$ scattering amplitude.}\label{tkdn}
\end{figure}

As an alternative possibility, the three body scattering in the $KDN$ system can also be formed as $K-(DN)_{\Lambda_c(2595)}$, 
with the $\Lambda _c(2595)$ assumed as a 
two body cluster. Now we need the two body matrix elements $KD$ and $KN$. The latter one is calculated using the chiral 
unitary approach that was mentioned before. The $KD$ matrix element, as also mentioned above, is investigated in \cite{danielfirst}. In order to calculate the 
form factor for the $\Lambda _c(2595)$, the cut off is needed. With the aforementioned strategy the value of  $\Lambda=973\mev$ is obtained and it was already used in section \ref{kbdn}. Using this cut off value, the $G_0(s)$ function is the same as in section \ref{kbdn} when we had scattering of a $\bar{K}$ on the cluster of $\Lambda_c(2595)$, since the masses of $\bar{K}$ and $K$ are the same. The $G_0(s)$ function is, thus the one plotted in Fig. \ref{fig3}.
Ultimately the total three-body scattering amplitude for the $K\Lambda_c(2595)$ scattering is evaluated and the result of $|T|^2$ is
 represented in Fig. \ref{tkdn}. There is an explicit narrow peak at $3100\mev$. 

One may try to investigate the structure of the peak in Fig. \ref{tkdn}, but it would only distract us from the main point, which is that the weight
of $|T|^2$ in Fig. \ref{tkdn} is very small compared with the one in Fig. \ref{tndk} for the $N-(DK)_{D_{s0}^*(2317)}$ configuration. Note that
a proper comparison requires to take into account the different field normalizations. Indeed, the S matrix for $N-(DK)_{D_{s0}^*(2317)}$ goes as

\begin{eqnarray}
S &\simeq& 1-i T 
\frac{1}{\sqrt{2\omega_{D_{s0}^*(2317)}}}
\frac{1}{\sqrt{2\omega'_{D_{s0}^*(2317)}}}
\sqrt{\frac{2M_N}{2E_N}}
\sqrt{\frac{2M_N}{2E'_N}},
\end{eqnarray}
while for $K-(DN)_{\Lambda_c(2595)}$ it goes as 

\begin{eqnarray}
S &\simeq& 1-i T 
\frac{1}{\sqrt{2\omega_K}}
\frac{1}{\sqrt{2\omega'_K}}
\sqrt{\frac{2M_{\Lambda_c}}{2E_{\Lambda_c}}}
\sqrt{\frac{2M_{\Lambda_c}}{2E'_{\Lambda_c}}}.
\end{eqnarray}
Hence, the proper comparison is $ \frac{T}{2m_{D_{s0}^*}}$ in the first case versus $ \frac{T}{2m_K}$ in the second, or $T(K(DN))$ versus
$ \frac{m_K}{m_{D_{s0}^*}}\times T(N(DK))$. Considering this, the strength of the peak for $|T(K(DN))|^{2}$  is about a factor $90$ smaller than for
$|T(N(DK))|^{2}$. This means that the $K(DN)$ configuration in the wave function of the $KDN$ system has a very small weight. Hence, we predict
a bound state of $NDK$ mostly made of a $N$ orbiting around a bound $DK$ cluster forming the ${D_{s0}^*(2317)}$.

\section{$ND \bar{D}$ interaction results}
\label{nddb}
\begin{figure}
\centering
\includegraphics[scale=1]{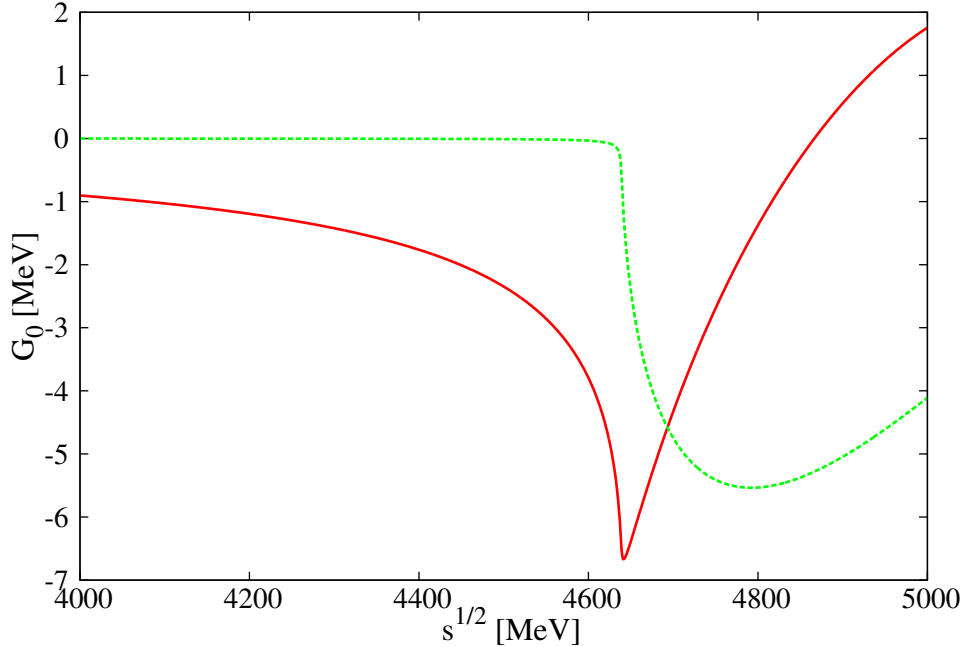}
\caption{Real(solid line) and imaginary(dashed line) parts of $G_0$ in $ND\bar{D}$.}\label{fig5}
\end{figure}

   The two-body $D\bar{D}$ interaction was investigated in \cite{danielfirst,Gamermann:2007mu,Molina:2008nh} and a resonance called $X(3700)$, was dynamically generated. This resonance would be the analogue one to the $f_0(980)$ which is basically a $K \bar{K}$ bound state \cite{Isgur,npa,norbert,prl,ramonet,hanhart1,hanhart2}. In our procedure we also reproduce this $D\bar{D}$ state successfully, getting the pole as $(3718+i0)\mev$ with the same parameters as in \cite{danielfirst}. We take a value of $\Lambda=850\mev$ from \cite{Gamermann:2007mu}, which is consistent with the methods mentioned above. Then we can calculate the form factor of the $X(3700)$ with Eq. \eqref{formfactor} by means of this cut off. Next we evaluate the $G_0$ by means of Eq. \eqref{g0meson}, for $N$ propagating between the $D$ and $\bar{D}$, and its results are shown in Fig. \ref{fig5}.

\begin{figure}
\centering
\includegraphics[scale=1]{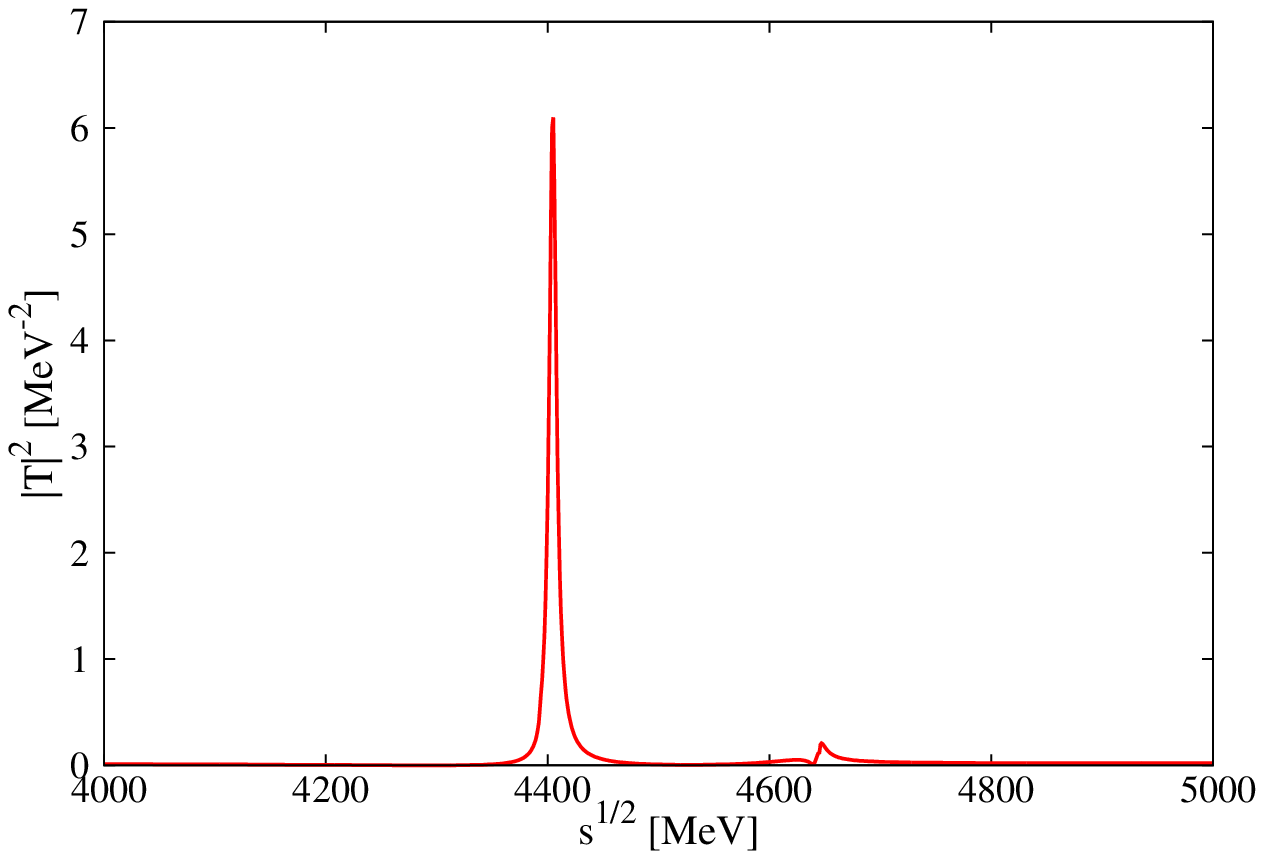}
\caption{Modulus squared of the $N X(3700)$ scattering amplitude.}\label{fig6}
\end{figure}

   The nucleon interaction with the $D,\bar{D}$ mesons was studied by the coupled channels two-body scattering equations in \cite{mizuangels,conlaura}. For the $DN$ scattering amplitude, as mentioned before, we followed the procedure of \cite{mizuangels}. For the $I=1$ sector there are eight coupled channels and we have used the same parameters as in $I=0$ which reproduced the $\Lambda_c(2595)$ resonance. The $\bar{D}N$ interaction, which is similar to the $KN$ channel \cite{mizuangels}, is repulsive in $I=1$ and vanishes for $I=0$. As in \cite{conlaura}, we took the same parameter as for the $DN$ interaction. Finally we obtain the $T$ matrix, for the $ND\bar{D}$ interaction by means Eq. \eqref{threet}, and show the results of $|T|^2$ in Fig. \ref{fig6}. From this figure we can see that there is a clear peak of $|T|^2$ around $4400\mev$ and the width is very small, less than $10\mev$. The peak appears below the $ND \bar{D}$ and $N X(3700)$ thresholds and corresponds to a bound state of $N X(3700)$. This would be a hidden charm baryon state of $J^P=\frac{1}{2}^+$ which appears in the same region of energies as other hidden charm states of $J^P=\frac{1}{2}^-$ obtained from the $\bar{D} \Lambda_c,~\bar{D} \Sigma_c$ and $\bar{D}^* \Lambda_c, ~\bar{D}^* \Sigma_c$ in \cite{wu,wudos}. In these latter works some reactions were suggested to observe those states in future Facilities. The same or similar reactions could be used to observe these states of positive parity.

\section{Conclusions}
We have investigated three body systems that have one $D$ meson or $D \bar{D}$, together with one baryon. The systems are $\bar{K}DN$, $NDK$ and $ND \bar{D}$. We have investigated the interaction of the systems taking into account that a couple of the particles in each case are strongly bound, forming dynamically generated states. We let then the third particle interact with the components of this cluster. Concretely, for $\bar{K}DN$ we study the scattering of $\bar{K}$ with the $D$ and $N$ components of the cluster of $DN$ that makes the $\Lambda_c(2595)$ resonance. In the second case, $NDK$, we find that the important configuration corresponds to $N$ scattering over the cluster of $KD$ that makes the $D_{s0}^*(2317)$. In the case of the  $ND\bar{D}$ we look at $N$ scattering on the $D \bar{D}$ cluster that is supposed to generate a bound state called X(3700), in analogy to the $f_0(980)$ which is mostly a bound state of $K \bar{K}$.  In all cases we find bound or quasibound states, relatively narrow, with energies $3150\mev, 3050\mev$ and $4400\mev$, respectively. All these states have $J^P=1/2^+$ and isospin $I=1/2$ and differ by their charm or strangeness content, $S=-1, C=1$, $S=1, C=1$, $S=0, C=0$, respectively.  The first state could perhaps be associated to the $\Xi(3123)$, which has unknown $J^P$, but the width obtained is a bit too large. The second state is of exotic nature and there is no counterpart in the PDG. The third state is a regular $N^*$ state as to quantum numbers, but it contains hidden charm. It lies in  an energy region where baryon states have not yet been investigated. We are thus making predictions in the frontier of this field and hope that with the coming Facilities of FAIR, or the BELLE upgrade, such states can be systematically studied and that our predictions can be confirmed.

\section*{Acknowledgments}
We thank J. Nieves and A. Ramos for helpful discussion. C. W. Xiao and M. Bayar thank J.J. Xie, R. Molina and E. J. Garzon for discussions and kind help. M. Bayar acknowledges support through the Scientific and Technical Research Council (TUBITAK)
BIDEP-2219 grant.
This work is partly supported by DGICYT contract  FIS2006-03438,
 the Generalitat Valenciana in the program Prometeo and 
the EU Integrated Infrastructure Initiative Hadron Physics
Project  under Grant Agreement n.227431.


\end{document}